\documentclass[pra,superscriptaddress]{revtex4}
\usepackage{graphicx}
\usepackage{dcolumn}
\usepackage{bm}
\usepackage{amsmath}
\usepackage{amssymb,amsbsy,cancel}
\usepackage{color,ulem}
\usepackage{soul}
\newcommand{\beq}{\begin{equation}}
\newcommand{\eeq}{\end{equation}}
\newcommand{\bea}{\begin{eqnarray}}
\newcommand{\eea}{\end{eqnarray}}
\newcommand{\bec}{\begin{center}}
\newcommand{\enc}{\end{center}}
\newcommand{\bfr}{\begin{flushright}}
\newcommand{\efr}{\end{flushright}}
\newcommand{\alp}{\alpha}

\newcommand{\om}{\omega}
\newcommand{\kap}{\kappa}
\newcommand{\gam}{\gamma}

\newcommand{\vp}{\varphi}

\newcommand{\tC}{\widetilde{C}}
\newcommand{\tL}{\widetilde{L}}

\newcommand{\tE}{\widetilde{E}}

\newcommand{\hb}{\hat{b}}

\newcommand{\hH}{\hat{H}}
\newcommand{\hsig}{\hat{\sigma}}
\begin{document}
\title{
Galvanically connected tunable coupler between a cavity and a waveguide
}
\author{Kazuki Koshino}
\email{kazuki.koshino@osamember.org}
\affiliation{College of Liberal Arts and Sciences, Tokyo Medical and Dental
University, Ichikawa, Chiba 272-0827, Japan}
\date{\today}
\begin{abstract}
One of the key technologies in recent quantum devices is 
the tunable coupling among quantum elements 
such as qubits, cavities, and waveguides. 
In this work, we propose a cavity-waveguide tunable coupler 
with an excellent on-off ratio, which is realized in 
a semi-infinite waveguide equipped with a tunable stub. 
The working principle of the present device is 
the shift of the node position of the cavity mode 
induced by the tunable boundary condition at the stub end. 
When the node position is adjusted to the branch point of the waveguide, 
the cavity mode becomes decoupled from the waveguide modes in principle. 
At the same time, owing to the galvanic connection, 
the present device readily achieves an ultrastrong cavity-waveguide coupling, 
where the cavity decay rate is comparable to the cavity resonance frequency. 
\end{abstract}
\maketitle

\section{introduction}
Regardless of their physical implementation, 
cavity quantum electrodynamics (QED) systems 
are commonly characterized by only several parameters, 
such as the resonance frequencies of the atom and the cavity ($\om_a$, $\om_c$), 
their mutual coupling rate ($g$), and their decay rates ($\gam$, $\kap$)~\cite{cqed1,cqed2}. 
One of the charms of cavity QED systems lies in their high designability. 
We can artificially set the cavity-related parameters ($\om_c$, $g$, and $\kap$)
through the design of the cavity.
In solid-state cavity QED systems using artificial atoms, 
the atom frequency $\om_a$ also becomes a designable parameter
and an unprecedentedly strong atom-cavity coupling $g$ becomes in reach~\cite{us1,us2}.

Cavity QED systems acquire further flexibility by the possibility of 
in-situ tuning of system parameters through the external fields. 
In circuit QED, a superconducting quantum interference device (SQUID) 
is used as a tunable element through the magnetic flux threading the loop~\cite{sq1}. 
For example, by replacing a Josephson junction composing a qubit with a SQUID, 
in-situ tuning of the qubit frequency becomes possible~\cite{ftq1,ftq2,ftq3}.
Such a frequency-tunable qubit is applicable to a tunable coupler between two qubits~\cite{qq0},  
which is indispensable to achieve a high two-qubit gate fidelity.  
Tunable couplers now play an essential role in constructing various quantum devices. 
Besides the qubit-qubit coupling~\cite{qq0,qq1,qq2,qq3,qq4,qq5,qq6}, 
tunable coupling has been developed in the cavity-cavity coupling~\cite{cc1,cc2,cc3}, 
the qubit-waveguide coupling~\cite{2012NJP,qw1,qw2,qw3}, 
and the cavity-waveguide coupling~\cite{cw0,cw1,cw2,cw3,cw4}.

In this study, we propose a cavity-waveguide tunable coupler
whose working principle differs fundamentally from the conventional tunable couplers. 
The proposed setup is a semi-infinite transmission line equipped with a tunable stub [Fig.~\ref{fig:sch}(a)], 
where the two finite ports (one infinite port) function as a cavity (waveguide). 
The cavity-waveguide coupling is tuned through the shift of the node position of the cavity mode. 
The cavity mode becomes completely decoupled from the waveguide modes in principle
when its node position is adjusted to the branch point of the waveguide. 
In contrast, due to the galvanic connection,
the cavity-waveguide coupling readily reaches the ultrastrong coupling regime,   
where the cavity decay rate amounts to the order of gigahertz, 
comparable to the resonance frequency.

The rest of this paper is organized as follows. 
In Sec.~\ref{sec:setup}, we present the setup investigated in this work, 
namely, a semi-infinite waveguide equipped with a tunable stub. 
In Sec.~\ref{sec:em}, we analyze the continuous eigenmodes of this waveguide. 
We observe the existence of a discrete cavity mode, 
which is 
decoupled from the propagating modes in the semi-infinite part of this waveguide, 
under a proper boundary condition at the stub end. 
In Sec.~\ref{sec:spec}, we analyze the microwave response of the cavity mode
to the stationary field input from the semi-infinite part. 
We focus on the phase shift of the input field upon reflection
and the photon energy stored in the cavity. 
From the results of microwave response, we determine in Sec.~\ref{sec:gtc} 
the resonance frequency and the linewidth of the cavity. 
We observe that the linewidth is extremely sensitive 
to the boundary condition at the stub end
and therefore that the cavity-waveguide coupling is widely tunable over several orders of magnitude.
We summarize this work in Sec.~\ref{sec:summary}.

\section{Setup}\label{sec:setup}
\begin{figure}
\begin{center}
\includegraphics[scale=1.0]{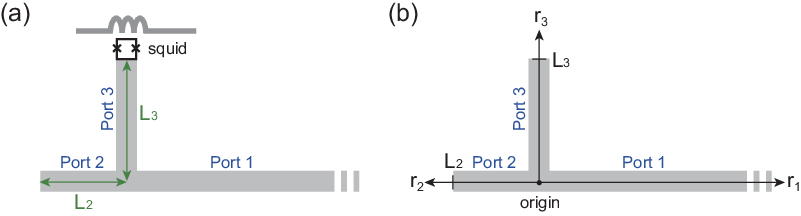}
\end{center}
\caption{(a)~Schematic of the investigated setup.
(b)~Coordinate system employed in this work.}
\label{fig:sch}
\end{figure}

In this study, we investigate a waveguide 
composed of three ports with the same property 
(characteristic impedance $Z$ and microwave phase velocity $v$),
as illustrated schematically in Fig.~\ref{fig:sch}(a).
Port~1 is semi-infinite, whereas Ports~2 and 3 
have finite lengths of $L_2$ and $L_3$, respectively. 
Port~2 is terminated by an infinitesimal capacitance to the ground
and the boundary condition there is open for the voltage. 
Port~3 is terminated by a SQUID so as to enable in-situ tuning of the boundary condition 
by the external magnetic flux threading the loop.
Setting the origin at the waveguide branch, 
we take a coordinate system depicted in Fig.~\ref{fig:sch}(b).
For concreteness, we employ the parameter values listed in Table~\ref{tab}.

\begin{table}
\caption{
List of parameters. $C_s$ and $E_s$ are the values 
for the two identical Josephson junctions forming the SQUID.
}
\label{tab}
\centering
\begin{tabular}{lll}
\hline
$v$ & (microwave velocity) & $10^8$~m/s
\\
$Z$ & (characteristic impedance) & $50~\Omega$
\\
$L_2$ & (length of Port~2) & 2.5~mm
\\
$L_3$ & (length of Port~3) & 4.5~mm
\\ 
$C_s$ & (capacitance) & 100~fF
\\
$(2e/\hbar)E_s$ & (critical current) & 5~$\mu$F 
\\
\hline
\end{tabular}
\end{table}

\section{Eigenmodes}\label{sec:em}

In this section, we investigate the eigenmodes of this waveguide. 
As a variable to describe the microwave propagating in this waveguide, 
we employ the flux (time-integrated voltage) defined by $\phi(r,t)=\int^t dt' V(r,t')$. 
Considering the semi-infinite nature of this waveguide, 
its eigenmodes are labelled by a continuous frequency $\om(>0)$.
The eigenmode function at frequency $\om$ is written as
\bea
\phi_{\om}(r) = 
\begin{cases}
\phi^{(1)}_{\om}(r_1) = 
\alp^{(1)}_{\om} \cos(\om r_1/v+\theta_{\om}) 
& (\mathrm{Port~1}) 
\\
\phi^{(2)}_{\om}(r_2) = \alp^{(2)}_{\om} \cos[\om(r_2-L_2)/v] 
& (\mathrm{Port~2}) 
\\
\phi^{(3)}_{\om}(r_3) = \alp^{(3)}_{\om} \cos[\om(r_3-L_{3,\om}^\mathrm{eff})/v] 
& (\mathrm{Port~3}) 
\end{cases},
\label{eq:fomr}
\eea
where $L_{3,\om}^\mathrm{eff}$ is the effective length of Port~3,
which is tunable through the magnetic flux threading the SQUID
(see Appendix~\ref{app:bc_squid}). 
$\theta_{\om}$ and the ratio of \{$\alp^{(1)}_{\om}$, 
$\alp^{(2)}_{\om}$, $\alp^{(3)}_{\om}$\}
are determined by the following 
boundary conditions at the waveguide branch (see Appendix~\ref{app:bc_branch}),
\bea
\phi^{(1)}_{\om}(0) = \phi^{(2)}_{\om}(0) = \phi^{(3)}_{\om}(0), \label{eq:bc1}
\\
\frac{d \phi^{(1)}_{\om}}{d r_1}(0) + 
\frac{d \phi^{(2)}_{\om}}{d r_2}(0) + 
\frac{d \phi^{(3)}_{\om}}{d r_3}(0) = 0. \label{eq:bc2}
\eea
Equations~(\ref{eq:bc1}) and (\ref{eq:bc2}) respectively represent 
the uniqueness of the voltage and the Kirchhoff's current law. 
From Eqs.~(\ref{eq:fomr})--(\ref{eq:bc2}), we have
\bea
\alp^{(1)}_{\om} \cos\theta_{\om} 
= \alp^{(2)}_{\om} \cos(L_2 \om/v) 
= \alp^{(3)}_{\om} \cos(L_{3,\om}^\mathrm{eff} \om/v),
\label{eq:cond1}
\\
\alp^{(1)}_{\om} \sin\theta_{\om} =
\alp^{(2)}_{\om} \sin(L_2 \om/v) + 
\alp^{(3)}_{\om} \sin(L_{3,\om}^\mathrm{eff} \om/v).
\label{eq:cond2}
\eea

\subsection{Special eigenmodes}
\begin{figure} 
\begin{center}
\includegraphics[width=130mm]{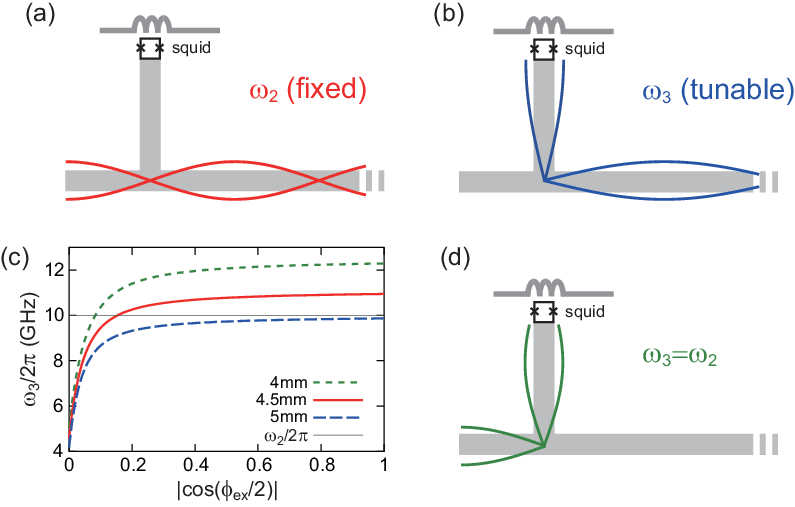}
\end{center}
\caption{
Eigenmode having a node at the waveguide branch. 
(a)~Eigenmode with vanishing amplitude in Port~3.
Its eigenfrequency is denoted by $\om_2$.
(b)~Eigenmode with vanishing amplitude in Port~2.
Its eigenfrequency is denoted by $\om_3$.
(c)~Tuning of $\om_3$ through the boundary condition. 
$\phi_\mathrm{ex}$ is the magnetic flux threading the SQUID loop
in units of the flux quantum.
Thin line plots $\om_2$, which is fixed at $2\pi\times 10$~GHz.
(d)~Cavity mode, the amplitude of which vanishes in Port~1.
This mode appears under a specific boundary condition, where $\om_3=\om_2$. 
}
\label{fig:spmodes}
\end{figure}

First, we consider the eigenmodes whose amplitudes vanish in Port~3.
Putting $\alp_{\om}^{(3)}=0$ in Eq.~(\ref{eq:cond1}), 
we observe that the eigenfrequencies of such modes satisfy $\cos(\om L_2/v)=0$.
Hereafter, we focus on the lowest eigenmode satisfying this condition. 
We define the frequency $\om_2$ by 
\bea
\om_2 L_2/v &=& \pi/2. \label{eq:k2L2}
\eea
At this frequency, we can confirm that 
$\alp^{(1)}_{\om_2}=\alp^{(2)}_{\om_2}$, 
$\alp^{(3)}_{\om_2}=0$, and $\theta_{\om_2}=\pi/2$. 
The spatial profile of this mode is schematically illustrated in Fig.~\ref{fig:spmodes}(a).
Similarly, we consider the lowest eigenmode whose amplitude vanishes in Port~2. 
The eigenfrequency $\om_3$ of this mode is determined by
\bea
\om_3 L_{3,\om_3}^\mathrm{eff}/v &=& \pi/2. \label{eq:k3L3}
\eea
Regarding this mode, we have 
$\alp^{(1)}_{\om_3}=\alp^{(3)}_{\om_3}$, 
$\alp^{(2)}_{\om_3}=0$, and $\theta_{\om_3}=\pi/2$. 
The spatial profile of this mode is schematically illustrated in Fig.~\ref{fig:spmodes}(b).

Note that $\om_2$ is a fixed value ($\om_2/2\pi=10$~GHz)
determined solely by $L_2$, 
whereas $\om_3$ is a tunable value through the boundary condition of Port~3 at the SQUID. 
In Fig.~\ref{fig:spmodes}(c), 
we show the dependence of $\om_3$ on the boundary condition
under the parameter values in Table~\ref{tab}. 
In the following part of this paper, we express 
the boundary condition at the end of Port~3 by the value of $\om_3$.
For $L_3=4.5$~mm, $\om_3/2\pi$ is tunable 
within the range from 4.567~GHz to 10.945~GHz.

\subsection{Cavity mode}
Next, we consider the eigenmodes whose amplitudes vanish in Port~1.
For these modes, the field amplitude is localized in a finite region, Ports~2 and 3. 
We refer to such localized modes as the {\it cavity} modes in this study. 
Putting $\alp_{\om}^{(1)}=0$ in Eq~(\ref{eq:cond1}), 
we immediately have $\cos(\om L_2/v)=0$ and $\cos(\om L_{3,\om}^\mathrm{eff}/v)=0$. 
This implies that 
such eigenmodes that are completely localized in a finite domain
can exist under a specific boundary condition at the SQUID.

Regarding the lowest cavity mode, 
the condition for the existence of a completely localized mode is
the exact tuning of $\om_3$ to $\om_2$. 
Its mode function is written as 
\bea
\phi_\mathrm{cav}(r) = \phi_0 \times
\begin{cases}
0 & (\mathrm{Port~1}) 
\\
- \sin(\om_2 r_2/v) & (\mathrm{Port~2}) 
\\
\sin(\om_2 r_3/v) & (\mathrm{Port~3}) 
\end{cases},
\label{eq:cavmode}
\eea
where $\phi_0$ is a constant. 
The spatial profile of this mode is schematically illustrated in Fig.~\ref{fig:spmodes}(d).

When $\om_3$ is exactly tuned to $\om_2$, 
the cavity mode is completely decoupled from the propagating modes in Port~1. 
In other words, the external decay rate $\kap$ of the cavity mode 
to the waveguide modes is zero in this case.
In contrast, when $\om_3$ is detuned slightly from $\om_2$, 
the cavity mode is weakly coupled from the propagating modes in Port~1
and $\kap$ takes a nonzero value. 
Then, the cavity mode becomes spectroscopically visible
by the input microwave applied from Port~1, 
as we discuss in Sec.~\ref{sec:spec}.

\subsection{General eigenmode}
For a general frequency [$\cos(\om L_2/v) \neq 0$ and $\cos(\om L_{3,\om}^\mathrm{eff}/v) \neq 0$], 
the eigenmode amplitudes do not vanish in all three ports.
From Eqs.~(\ref{eq:cond1})--(\ref{eq:cond2}), 
$\theta_{\om}$, $\alp^{(2)}_{\om}/\alp^{(1)}_{\om}$ and $\alp^{(3)}_{\om}/\alp^{(1)}_{\om}$ 
are determined by the following equations,
\bea
\tan \theta_{\om} &=& \tan(\om L_2/v) + \tan(\om L_{3,\om}^\mathrm{eff}/v),
\label{eq:vp}
\\
\frac{\alp^{(2)}_{\om}}{\alp^{(1)}_{\om}} 
&=& 
\frac{\cos\theta_{\om}}{\cos(\om L_2/v)},
\\
\frac{\alp^{(3)}_{\om}}{\alp^{(1)}_{\om}}  
&=& 
\frac{\cos\theta_{\om}}{\cos(\om L_{3,\om}^\mathrm{eff}/v)}.
\eea

\section{Spectroscopy of cavity mode}\label{sec:spec}

\subsection{Phase shift upon reflection}
\begin{figure} 
\begin{center}
\includegraphics[width=140mm]{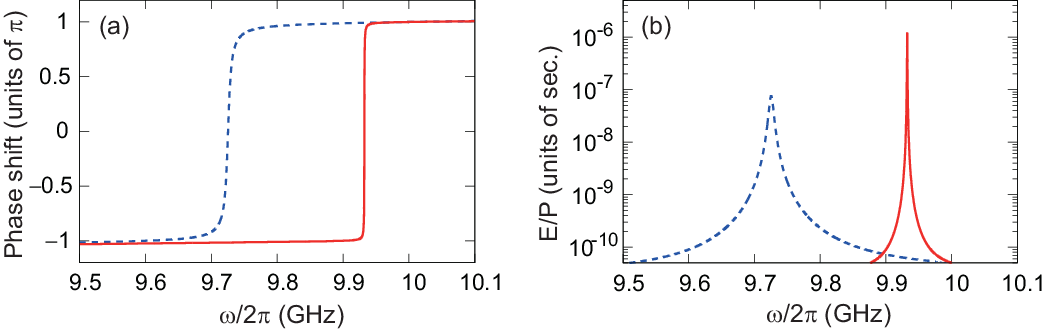}
\end{center}
\caption{
Spectroscopy of cavity mode. 
(a)~Phase shift of the input field from Port~1 upon reflection and 
(b)~cavity photon energy normalized by the input power, 
plotted against the input frequency $\om$.
The boundary condition at the SQUID is set so that  
$\om_3/2\pi=9.6$~GHz (blue dotted) and 9.9~GHz (red solid).
}
\label{fig:effcav}
\end{figure}

Under a general boundary condition at the SQUID (where $\om_3 \neq \om_2$), 
the cavity mode (Ports~2 and 3) is coupled to the waveguide modes (Port~1) 
and responds to a microwave signal input through Port~1.
In this subsection, 
we investigate the phase shift upon reflection 
of a stationary input field.
From the eigenmode function [Eq.~(\ref{eq:fomr})] in Port~1, 
this phase shift is identified as $2\theta_{\om}$,
where $\theta_{\om}$ is determined by Eq.~(\ref{eq:vp}).
This is plotted against the input frequency $\om$ in Fig.~\ref{fig:effcav}(a),
varying the boundary condition at the SQUID.
We observe an abrupt increase of the phase shift by $2\pi$ 
around a certain frequency $\om_c$ and within a certain bandwidth $\kap$. 
This fact supports that 
Ports~2 and 3 function as an effective cavity mode
with the central frequency $\om_c$ and the linewidth $\kap$. 
We also observe that 
$\om_c$ and $\kap$ are sensitive to the boundary condition, 
as we will discuss in detail in Sec.~\ref{sec:gtc}.

\subsection{Cavity photon energy}
In this subsection, 
we investigate the photon energy stored in the cavity mode. 
We consider a stationary field at frequency $\om$
whose waveform is given by $\phi(r,t)=\phi_{\om}(r) e^{-i\om t}$, 
where 
$\phi_{\om}(r)$ is the eigenmode function [Eq.~(\ref{eq:fomr})] at frequency $\om$.
The energy density $\tE$ per unit length of the waveguide is written as
\bea
\tE =
\frac{\tC}{2}\left|\frac{\partial \phi}{\partial t}\right|^2 
+ \frac{1}{2\tL}\left|\frac{\partial \phi}{\partial r}\right|^2, 
\label{eq:tE}
\eea
where $\tC$ and $\tL$ respectively denote 
the capacitance and inductance per unit length, 
which are related to the microwave velocity $v$ 
and the characteristic impedance $Z$ of this waveguide
by $\tC=1/(vZ)$ and $\tL=Z/v$.
Integrating the energy density $\tE$ in the cavity part (Ports~2 and 3), 
the photon energy $E$ stored in the cavity is given by
\bea
E = \frac{\om^2}{2vZ} 
\left[(\alp^{(2)}_{\om})^2 L_2 + (\alp^{(3)}_{\om})^2 L_3 \right].
\label{eq:cavene}
\eea
Regarding the input field propagating in Port~1,
from Eq.~(\ref{eq:fomr}), it is identified as
$(\alp_{\om}^{(1)}/2)\times e^{-i[\om(r_1/v+t)+\theta_{\om}]}$.
Therefore, the power $P(=v\tE)$ of the input field is given by
$P = {\om^2 (\alp^{(1)}_{\om})^2}/{4Z}$.
The cavity photon energy normalized by the input power is given by
\bea
E/P = \frac{2}{v}
\left[(\alp^{(2)}_{\om}/\alp^{(1)}_{\om})^2 L_2 + (\alp^{(3)}_{\om}/\alp^{(1)}_{\om})^2 L_3 \right], 
\label{eq:EP1}
\eea
which depends only on the input frequency $\om$
and is insensitive to the field strength. 
In Fig.~\ref{fig:effcav}(b), we plot $E/P$ evaluated 
by Eq.~(\ref{eq:EP1}) against the input frequency $\om$. 
We observe a sharp peak around a certain frequency $\om_c$. 
This fact also supports that 
Ports~2 and 3 function as an effective cavity mode.

On the other hand, the standard quantum-optics theory predicts that,
for a cavity with the central frequency $\om_c$ and the linewidth $\kap$,
$E/P$ has a Lorentzian shape as given by
\bea
E/P &=& \frac{\kap}{(\om-\om_c)^2 + \kap^2/4}.
\label{eq:EP2}
\eea
We can confirm that the lineshape of $E/P$ 
is a Lorentzian in agreement with Eq.~(\ref{eq:EP2}).

\section{Tuning of cavity parameters}\label{sec:gtc}
\subsection{Determination of cavity parameters}
We can identify the resonance frequency $\om_c$ and the linewidth $\kap$ of the cavity mode 
from the phase shift of a stationary input field upon reflection [Fig.~\ref{fig:effcav}(a)]. 
$\om_c$ is identified as the frequency at which the phase shift becomes zero, 
whereas $\kap$ is identified as the difference in frequencies at which the phase shift becomes $\pm \pi/2$.
Alternatively, we can  determine $\om_c$ and $\kap$
from the cavity photon energy normalized by the input power [Fig.~\ref{fig:effcav}(b)].
$\om_c$ and $\kap$ are identified as the peak position and the linewidth of 
the Lorentzian, respectively. 
The resonance frequency $\om_c$ and the linewidth $\kap$ 
thus determined are respectively plotted in Figs.~\ref{fig:gtc}(a) and (b), 
varying the boundary condition. 
We observe that the above two methods yield almost identical results. 

\subsection{Dependence of cavity parameters on boundary condition}
\begin{figure} 
\begin{center}
\includegraphics[width=140mm]{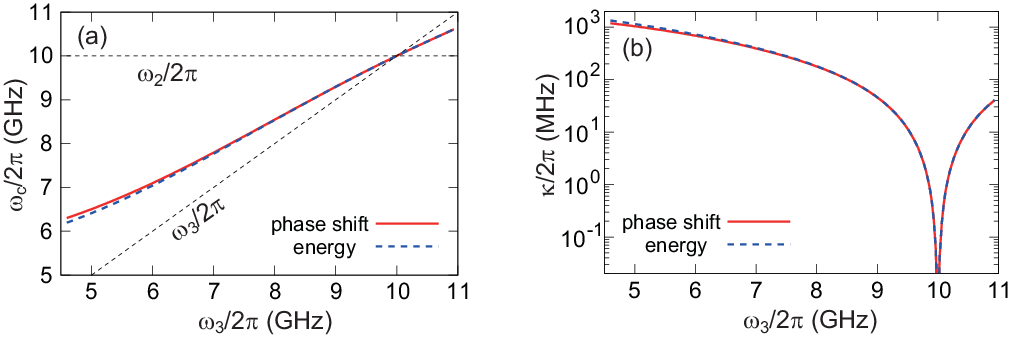}
\end{center}
\caption{
Dependences of (a)~central frequency $\om_c$ and (b)~linewidth $\kap$ of the cavity mode, 
on the boundary condition at the SQUID. 
Red solid (blue dotted) lines plot the values 
estimated from the phase shift upon reflection (the cavity photon energy).
In (a), $\om_2$ (fixed at $2\pi\times 10$~GHz) 
and $\om_3$ are also plotted by thin lines for reference.
}
\label{fig:gtc}
\end{figure}

As we observe in Fig.~\ref{fig:gtc}(a),
the resonance frequency $\om_c$ lies between $\om_2$ and $\om_3$
and exhibits an almost linear dependence on $\om_3$.
In contrast, as we observe in Fig.~\ref{fig:gtc}(b), 
the linewidth $\kap$ depends drastically on the boundary condition. 
In particular, when $\om_3$ is tuned exactly to $\om_2$, 
the linewidth $\kap$ vanishes in principle.
In this case, the cavity mode extending in Ports~2 and 3 
has a node at the waveguide branch [Fig.~\ref{fig:spmodes}(d)]
and becomes completely decoupled from the propagating modes in Port~1. 
If the boundary condition is slightly varied from this state,  
the node position is shifted from the waveguide branch
and coupling to the propagating modes in Port~1 is recovered. 
%
%

When the detuning between $\om_3$ and $\om_2$ is large, 
in clear contrast with the case of small detuning, 
the cavity-waveguide coupling $\kap$ readily reach the order of a gigahertz. 
This is because our setup contains no circuit element 
such as a capacitance
that clearly divides the cavity and the waveguide 
and sets the upper limit on their coupling. 
Thus, in the present device, 
a high on-off ratio of the cavity-waveguide coupling is expected.

\subsection{Critical photon number}
\label{ssec:cpn}
In derivation of the boundary condition at the SQUID (Appendix~\ref{app:bc_squid}), 
we employ a linear approximation [$\sin(2e\phi/\hbar) \approx 2e\phi/\hbar$]  
to the flux field at the SQUID position. 
This requires that
the flux there is sufficiently smaller than the magnetic flux quantum ($\hbar/2e$)
and sets a critical photon number $N_\mathrm{crit}$ to the cavity,  
above which the nonlinearity of this cavity gradually becomes apparent.

Considering the flux at the SQUID position 
[$r_3=L_3$ in Eq.~(\ref{eq:cavmode})], 
the condition for the linealization is written as
\bea
\left| \phi_0 \sin[\pi L_3/(2L_2)] \right|
\lesssim 
\hbar/2e.
\label{eq:cond_lin}
\eea
On the other hand, 
integrating Eq.~(\ref{eq:tE}) in Ports~2 and 3 and using $N=E/(\hbar\om_2)$, 
the cavity photon number $N$ is given by
\bea
N = \frac{\pi(1+L_3/L_2)\phi_0^2}{4\hbar Z}.
\label{eq:N}
\eea
From Eqs.~(\ref{eq:cond_lin}) and (\ref{eq:N}), 
the critical photon number is estimated to be
\bea
N_\mathrm{crit} 
\sim
\frac{\pi\hbar(1+L_3/L_2)}{16 e^2 Z \sin^2(\pi L_3/2L_2)}. 
\eea

\begin{figure} 
\begin{center}
\includegraphics[width=70mm]{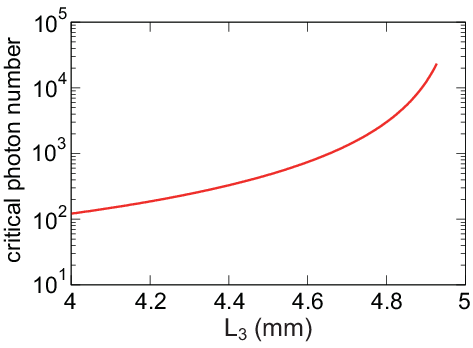}
\end{center}
\caption{
Critical photon number for the lowest cavity mode at 10~GHz. 
Note that tuning of $\om_3/2\pi$ to 10~GHz is impossible 
under any boundary condition at the SQUID for $L_3 > 4.93$~mm. 
}
\label{fig:Ncrit}
\end{figure}

In Fig.~\ref{fig:Ncrit}, we plot the critical photon number 
of the lowest cavity mode, varying the length $L_3$ of Port~3. 
%
%
The cavity mode amplitude at the SQUID position
is proportional to $\sin(\om_2 L_3/v)$ [Eq.~(\ref{eq:cavmode})]
and becomes smaller as $L_3$ approaches to 5~mm ($=\pi v/\om_2$). 
As a result, the critical photon number increases in this limit. 
However, note that we cannot tune $\om_3$ to $\om_2$ 
for $L_3>4.93$~mm, as we observe in Fig.~\ref{fig:spmodes}(c). 

\section{summary}\label{sec:summary}
In this study, we theoretically propose 
a galvanically connected cavity-waveguide tunable coupler. 
The investigated setup is a waveguide 
composed of three ports with the same property:
one port is semi-infinite, whereas the other two ports have finite lengths. 
One of the finite ports is terminated by a SQUID and functions as a tunable stub.
We analyzed the microwave response of this waveguide
using its continuous eigenmodes 
and observed that this setup functions as 
a tunable cavity-waveguide tunable coupler 
under adequate choice of the lengths of the finite ports.
The working principle of this tunable coupler
is the shift of the node position of the cavity mode 
with respect to the waveguide branch. 
Due to the galvanic connection, this device enables 
an excellent on-off ratio in the cavity-waveguide coupling, 
which is applicable to the generation of an ultrashort microwave pulse for example.

\section*{Acknowledgments}
The author is grateful to K. Mizuno, Y. Sunada, T. Yamamoto, 
and K. Semba for fruitful discussions.
This work was supported by JST Moonshot R\&D (JPMJMS2061, JPMJMS2067) 
and JSPS KAKENHI (22K03494).

\appendix

\section{Boundary condition at SQUID}
\label{app:bc_squid}
As the SQUID terminating Port~3, we consider the one 
composed of two identical Josephson junctions 
(each having capacitance $C_s$ and Josephson energy $E_s$) forming a loop. 
We denote the external magnetic flux threading the loop 
by $(\hbar/2e)\phi_\mathrm{ex}$. 
Then, after linearization [$\sin(2e\phi/\hbar) \approx2e\phi/\hbar $],
the boundary condition at the SQUID position is written as~\cite{2012NJP}
\bea
\tC_s \frac{\partial^2 \phi}{\partial t^2}=
-\left(\frac{2e}{\hbar}\right)^2\tE_s(\phi_\mathrm{ex})\phi
-\frac{1}{\tL}\frac{\partial \phi}{\partial r},
\label{eq:bcsquid}
\eea
where $\tC_s=2C_s$, $\tE_s(\phi_\mathrm{ex})=2E_s|\cos(\phi_\mathrm{ex}/2)|$, 
and $\tL(=Z/v)$ is the inductance of the waveguide per unit length.
Putting $\phi(r,t)=\phi_{\om}^{(3)}(r_3) e^{-i\om t}$ in Eq.~(\ref{eq:bcsquid}),
where $\phi_{\om}^{(3)}(r_3)$ is given by Eq.~(\ref{eq:fomr}), we obtain
\bea
\tan[\om(L_{3,\om}^\mathrm{eff}-L_3)/v] = 
2ZC_s\om - \frac{8e^2ZE_s}{\hbar^2\om}|\cos(\phi_\mathrm{ex}/2)|.
\label{eq:tan}
\eea
This is an equation to determine $L_{3,\om}^\mathrm{eff}$ for a given frequency $\om$.

Putting $\om=\om_3$ in Eq.~(\ref{eq:tan}) and using Eq.~(\ref{eq:k3L3}), we obtain 
\bea
\cot(\om_3L_3/v) = 2ZC_s\om_3 - \frac{8e^2ZE_s}{\hbar^2\om_3}|\cos(\phi_\mathrm{ex}/2)|.
\label{eq:cot}
\eea
This is an equation to determine $\om_3$. 
The numerical solution of this equation is shown 
in Fig.~\ref{fig:spmodes}(c) in the main text. 

\section{Boundary condition at waveguide branch}
\label{app:bc_branch}
\begin{figure}[h]
\begin{center}
\includegraphics[width=70mm]{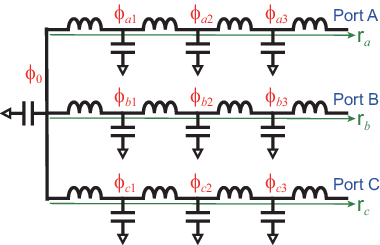}
\end{center}
\caption{
Circuit diagram of a waveguide branch.
All capacitors (inductors) have infinitesimal capacitance $\Delta C$
(inductance $\Delta L$). 
}
\label{fig:bra}
\end{figure}
Here, we derive the boundary condition at a waveguide branch
from the circuit model having three ports A, B, and C (Fig.~\ref{fig:bra}).
The classical Lagrangian describing this circuit is given by
\bea
L &=& \frac{\Delta C}{2}\dot{\phi}_0^2 -\frac{1}{2\Delta L} 
\left[ (\phi_0-\phi_{a1})^2 + (\phi_0-\phi_{b1})^2 + (\phi_0-\phi_{c1})^2\right]
\nonumber \\
&+& \frac{\Delta C}{2}\left[ \dot{\phi}_{a1}^2 + \dot{\phi}_{b1}^2 + \dot{\phi}_{c1}^2\right]
-\frac{1}{2\Delta L} 
\left[ (\phi_{a1}-\phi_{a2})^2 + (\phi_{b1}-\phi_{b2})^2 + (\phi_{c1}-\phi_{c2})^2\right] 
\nonumber \\
&+& \cdots. 
\eea
From this Lagrangian, we can derive
the equation of motion for the flux $\phi_0$ at the branch point,
\bea
\Delta C \ddot{\phi}_0 = [(\phi_{a1}-\phi_0)+(\phi_{b1}-\phi_0)+(\phi_{c1}-\phi_0)]/\Delta L.
\label{eq:eqmphi0}
\eea
We here switch to the continuous description of the flux field,
namely, $\phi_{aj}(t)=\phi_a(j\Delta r, t)$, 
where $\Delta r$ is the infinitesimal distance between the nodes.
$\phi_b(r_b,t)$ and $\phi_c(r_c,t)$ are introduced similarly.
Since the flux $\phi_0$ is common to the three semi-infinite waveguides, we immediately have
\bea
\phi_a(0,t) = \phi_b(0,t)=\phi_c(0,t).
\label{eq:bc1app}
\eea
With the continuous description, Eq.~(\ref{eq:eqmphi0}) is rewritten as
\bea
\Delta C \frac{\partial^2 \phi_a}{\partial t^2}(0,t) =
\frac{1}{\tL}\left[\frac{\partial \phi_a}{\partial r_a}(0,t) + 
\frac{\partial \phi_b}{\partial r_b}(0,t) + \frac{\partial \phi_c}{\partial r_c}(0,t)
\right],
\label{eq:eqmphi1}
\eea
where $\tL=\Delta L/\Delta r$ is the inductance per unit length. 
Since the left-hand side of Eq.~(\ref{eq:eqmphi1}) is proportional to $\Delta C$ 
and is therefore infinitesimal, we obtain
\bea
\frac{\partial \phi_a}{\partial r_a}(0,t) + \frac{\partial \phi_b}{\partial r_b}(0,t) 
+ \frac{\partial \phi_c}{\partial r_c}(0,t) = 0.
\label{eq:bc2app}
\eea
Equations~(\ref{eq:bc1app}) and (\ref{eq:bc2app}) are the boundary conditions at the waveguide branch.



\begin{thebibliography}{99}
\bibitem{cqed1}
H. Walther, B. T. H. Varcoe, B.-G. Englert, and T. Becker, 
{\it Cavity quantum electrodynamics}, 
Rep. Prog. Phys. {\bf 69}, 1325 (2006).
\bibitem{cqed2}
A. Blais, A. L. Grimsmo, S. M. Girvin, and A. Wallraff, 
{\it Circuit quantum electrodynamics}, 
Rev. Mod. Phys. {\bf 93}, 025005 (2021).

\bibitem{us1}
A. F. Kockum, A. Miranowicz, S. De Liberato, S. Savasta, and F. Nori, 
{\it Ultrastrong coupling between light and matter}, 
Nat. Rev. Phys. {\bf 1}, 19 (2019).
\bibitem{us2}
P. Forn-Diaz, L. Lamata, E. Rico, J. Kono, and E. Solano, 
{\it Ultrastrong coupling regimes of light-matter interaction}, 
Rev. Mod. Phys. {\bf 91}, 025005 (2019).

\bibitem{sq1}
M. Wallquist, V. S. Shumeiko, and G. Wendin, 
{\it Selective coupling of superconducting charge qubits mediated by a tunable stripline cavity},
Phys. Rev. B {\bf 74}, 224506 (2006). 


\bibitem{ftq1}
L. DiCarlo, J. M. Chow, J. M. Gambetta, L. S. Bishop, B. R. Johnson, D. I. Schuster, 
J. Majer, A. Blais, L. Frunzio, S. M. Girvin, and R. J. Schoelkopf,
{\it Demonstration of two-qubit algorithms with a superconducting quantum processor}, 
Nature {\bf 460}, 240 (2009).
%
\bibitem{ftq2}
R. Barends, J. Kelly, A. Megrant, A. Veitia, D. Sank, E. Jeffrey, T. C. White, J. Mutus, 
A. G. Fowler, B. Campbell, Y. Chen, Z. Chen, B. Chiaro, A. Dunsworth, C. Neill, 
P. O'Malley, P. Roushan, A. Vainsencher, J. Wenner, A. N. Korotkov, A. N. Cleland, and J. M. Martinis,
{\it Superconducting quantum circuits at the surface code threshold for fault tolerance}, 
Nature {\bf 508}, 500 (2014).
%
\bibitem{ftq3}
J. Kelly, R. Barends, A. G. Fowler, A. Megrant, E. Jeffrey, T. C. White, D. Sank, 
J. Y. Mutus, B. Campbell, Y. Chen, Z. Chen, B. Chiaro, A. Dunsworth, I.-C. Hoi, 
C. Neill, P. J. J. O'Malley, C. Quintana, P. Roushan, A. Vainsencher, 
J. Wenner, A. N. Cleland, and J. M. Martinis,
{\it State preservation by repetitive error detection in a superconducting quantum circuit},
Nature {\bf 519}, 66 (2015).

\bibitem{qq0}
F. Yan, P. Krantz, Y. Sung, M. Kjaergaard, D. L. Campbell, T. P. Orlando, 
S. Gustavsson, and W. D. Oliver, 
{\it Tunable Coupling Scheme for Implementing High-Fidelity Two-Qubit Gates},
Phys. Rev. Appl. {\bf 10}, 054062 (2018).
%
\bibitem{qq1}
A. O. Niskanen, K. Harrabi, F. Yoshihara, Y. Nakamura, S. Lloyd, and J. S. Tsai, 
{\it Quantum Coherent Tunable Coupling of Superconducting Qubits}, 
Science {\bf 316}, 723 (2007).
%
\bibitem{qq2}
Y. Chen, C. Neill, P. Roushan, N. Leung, M. Fang, R. Barends, J. Kelly, 
B. Campbell, Z. Chen, B. Chiaro, A. Dunsworth, E. Jeffrey, A. Megrant, 
J. Y. Mutus, P. J. J. O'Malley, C. M. Quintana, D. Sank, A. Vainsencher, 
J. Wenner, T. C. White, Michael R. Geller, A. N. Cleland, and J. M. Martinis, 
{\it Qubit Architecture with High Coherence and Fast Tunable Coupling},
Phys. Rev. Lett. {\bf 113}, 220502 (2014).
%
\bibitem{qq3}
M. Kounalakis, C. Dickel, A. Bruno,N. K. Langford, and G. A. Steele,  
{\it Tuneable hopping and nonlinear cross-Kerr interactions in a high-coherence superconducting circuit}, 
npj Quantum Information {\bf 4}, 38 (2018). 
%
\bibitem{qq4}
H. Goto,
{\it Double-Transmon Coupler: Fast Two-Qubit Gate with No Residual Coupling for Highly Detuned Superconducting Qubits}, 
Phys. Rev. Appl. {\bf 18}, 034038 (2022). 
%
\bibitem{qq5}
D. L. Campbell, A. Kamal, L. Ranzani, M. Senatore, and M. D. LaHaye,
{\it Modular Tunable Coupler for Superconducting Circuits},
Phys. Rev. Applied {\bf 19}, 064043 (2023).
%
\bibitem{qq6}
H. Zhang, C. Ding, D. K. Weiss, Z. Huang, Y. Ma, C. Guinn, S. Sussman, 
S. P. Chitta, D. Chen, A. A. Houck, J. Koch, and D. I. Schuster, 
{\it Tunable Inductive Coupler for High-Fidelity Gates Between Fluxonium Qubits},
PRX Quantum {\bf 5}, 020326 (2024)


\bibitem{cc1}
A. Baust, E. Hoffmann, M. Haeberlein, M. J. Schwarz, P. Eder, J. Goetz, 
F. Wulschner, E. Xie, L. Zhong, F. Quijandria, B. Peropadre, D. Zueco, 
J.-J. Garcia Ripoll, E. Solano, K. Fedorov, E. P. Menzel, F. Deppe, 
A. Marx, and R. Gross
{\it Tunable and switchable coupling between two superconducting resonators},
Phys. Rev. B {\bf 91}, 014515 (2015).
%
\bibitem{cc2}
F. Wulschner, J. Goetz, F. R. Koessel, E. Hoffmann, A. Baust, P. Eder, 
M. Fischer, M. Haeberlein, M. J. Schwarz, M. Pernpeintner, E. Xie, 
L. Zhong, C. W. Zollitsch, B. Peropadre, J. J. G. Ripoll, E. Solano, 
K. G. Fedorov, E. P. Menzel, F. Deppe, A. Marx, and R. Gross, 
{\it Tunable coupling of transmission-line microwave resonators mediated by an rf SQUID}, 
EPJ Quantum Technol. {\bf 3}, 10 (2016). 
%
\bibitem{cc3}
T. Miyanaga, A. Tomonaga, H. Ito, H. Mukai, and J.S. Tsai, 
{\it Ultrastrong Tunable Coupler Between Superconducting LC Resonators}, 
Phys. Rev. Appl. {\bf 16}, 064041 (2021).

\bibitem{2012NJP}
K. Koshino and Y. Nakamura,
{\it Control of the radiative level shift and linewidth of a superconducting 
artificial atom through a variable boundary condition},
New J. Phys. {\bf 14}, 043005 (2012). 
%
\bibitem{qw1}
P. Forn-Diaz, C. W. Warren, C. W. S. Chang, A. M. Vadiraj, and C. M. Wilson, 
{\it On-Demand Microwave Generator of Shaped Single Photons},
Phys. Rev. Appl.{\bf 8}, 054015 (2017).
%
\bibitem{qw2}
P. Kurpiers, P. Magnard, T. Walter, B. Royer, M. Pechal, J. Heinsoo, Y. Salathe, A. Akin, S. Storz, J.-C. Besse, S. Gasparinetti, A. Blais, and A. Wallraff, 
{\it Deterministic quantum state transfer and remote entanglement using microwave photons},
Nature {\bf 558}, 264 (2018).
%
\bibitem{qw3}
N. Janzen, X. Dai, S. Ren, J. Shi, and A. Lupascu, 
{\it Tunable coupler for mediating interactions between a two-level system 
and a waveguide from a decoupled state to the ultrastrong coupling regime},
Phys. Rev. Res.{\bf 5}, 033155 (2023).

\bibitem{cw0}
Yi Yin, Yu Chen, Daniel Sank, P. J. J. O'Malley, T. C. White, R. Barends, J. Kelly, 
E. Lucero, M. Mariantoni, A. Megrant, C. Neill, A. Vainsencher, 
J. Wenner, A. N. Korotkov, A. N. Cleland, and J. M. Martinis, 
{\it Catch and Release of Microwave Photon States}, 
Phys. Rev. Lett. {\bf 110}, 107001 (2013).

\bibitem{cw1}
E. A. Sete, A. Galiautdinov, E. Mlinar, J. M. Martinis, and A. N. Korotkov, 
{\it Catch-Disperse-Release Readout for Superconducting Qubits},
Phys. Rev. Lett. {\bf 110}, 210501 (2013).

\bibitem{cw2}
M. Pierre, I.-M. Svensson, S. R. Sathyamoorthy, G. Johansson, P. Delsing, 
{\it Storage and on-demand release of microwaves using superconducting 
resonators with tunable coupling}, 
Appl. Phys. Lett. {\bf 104}, 232604 (2014). 

\bibitem{cw3}
C. Bockstiegel, Y. Wang, M. R. Vissers, L. F. Wei, S. Chaudhuri, J. Hubmayr, and J. Gao, 
{\it A tunable coupler for superconducting microwave resonators 
using a nonlinear kinetic inductance transmission line},
Appl. Phys. Lett. {\bf 108}, 222604 (2016). 

\bibitem{cw4}
M. Pierre, S. R. Sathyamoorthy, I.-M. Svensson, G. Johansson, and P. Delsing, 
{\it Resonant and off-resonant microwave signal manipulation 
in coupled superconducting resonators}, 
Phys. Rev. B {\bf 99}, 094518 (2019). 

\end{thebibliography}
\end{document}